\documentclass[12pt]{article}
\input epsf.tex 
\def\beq{\begin{equation}} 
\def\eeq{\end{equation}} 
\def\eeq{\end{equation}} 
\def\bea{\begin{eqnarray}} 
\def\eea{\end{eqnarray}} 

\newcommand{\gsim}{\stackrel{>}{_\sim}}
\newcommand{\lsim}{\stackrel{<}{_\sim}}
\title{The Little Higgs from a Simple Group}
 
\author{
        D.E. Kaplan\,$^{a}$\thanks{\tt dkaplan@pha.jhu.edu}\ \
        and M. Schmaltz\,$^b$\thanks{\tt schmaltz@bu.edu} \ \ \\ \\ 
        \small \sl $^a$\ Department of Physics and Astronomy,  \\
        \small \sl Johns Hopkins University, Baltimore, MD  21218\\
        \small \sl $^b$\ Physics Department, Boston University, 
                Boston, MA  02215\\ \\ 
       }

\begin{document} 
\baselineskip=17pt 
\pagestyle{plain}

\begin{titlepage} 
\vskip-.4in 
\maketitle 
\begin{picture}(0,0)(0,0) 
\put(308,350){BUHEP-03-03}
\end{picture}

\begin{abstract} 
\leftskip-.6in 
\rightskip-.6in 
\vskip.4in 

We present a model of electroweak symmetry breaking in which the Higgs boson is
a pseudo-Nambu-Goldstone boson.  By embedding the standard models 
$SU(2)\times U(1)$ into an $SU(4)\times U(1)$ gauge group, one-loop
quadratic divergences to the Higgs mass from gauge and top loops are
canceled automatically with the minimal particle content.
The potential contains a Higgs quartic coupling which does not introduce
one-loop quadratic divergences.
Our theory is weakly coupled at the electroweak scale, it has new
weakly coupled particles at the TeV scale and a cutoff above 10 TeV,
all without fine tuning. We discuss the spectrum of the model and
estimate the constraints from electroweak precision measurements. 

\end{abstract} 
\thispagestyle{empty} 
\setcounter{page}{0} 
\end{titlepage} 
 
{\it ``He who hath clean hands and a good heart is okay in my
        book, but he who fools around with barnyard animals has
        got to be watched''}\\
\phantom{xxxxxxxxxxxxxxxxxxxxxxxxxxxxxxxxxxxxxxxxxxxxxxxxxxxxxx} - W. Allen
\vspace{.5cm}

{\it ``The littler the Higgs the bigger the group''}\\
\phantom{xxxxxxxxxxxxxxxxxxxxxxxxxxxxxxxxxxxxxxxxxxxxxxxxxxxx} - S. Glashow
\vspace{1.0cm}
 
\section{Introduction} 

The Standard Model (SM) is well supported by all high
energy data \cite{PDG}. Precision tests match predictions
including one-loop quantum corrections. This suggest
the SM is a valid description of Nature up to energies in
the multi-TeV range with a Higgs mass which is less than
about 200 GeV \cite{LEPSLC}.

This picture is not satisfying because the Higgs gets one-loop quadratically
divergent corrections to its squared mass.  The most significant corrections
come from loops of top quarks, $W$-bosons and the Higgs (Figure 1).
The top loop is the most severe - demanding a contribution to the Higgs
mass of 200 GeV or less requires a momentum cutoff
$\Lambda_{top} \lsim 700$ GeV.
If one allows fine tuning of order 10\% between this correction and a
counter term, one still needs $\Lambda_{top} \lsim 2$ TeV.

\begin{figure}[htb]
\vskip 0.0truein
\centerline{\epsfysize=1.6in
{\epsffile{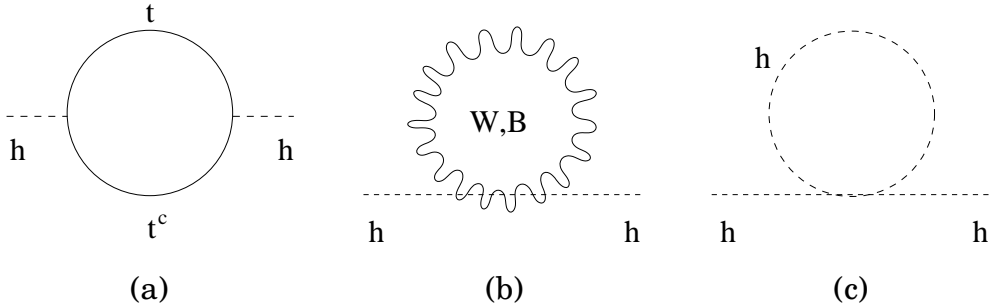}}}
\vskip 0.0truein
\caption[]{\it Quadratically divergent contributions to the Higgs mass
in the Standard model: top loop, $SU(2)\times U(1)$ gauge boson loops
and Higgs loop. }
\label{fig:loop}
\end{figure}

The need to cancel quadratic divergences along with the consistency of 
electroweak precision measurements with the SM suggests new 
{\it weakly coupled} physics at $\sim$ 1 TeV.

Supersymmetry softly broken around 1 TeV is an example of new physics that 
meets these criteria.  Loops of superpartners cancel all quadratic divergences 
in the SM.  The minimal supersymmetric standard model's most
compelling feature is its suggestive unification of couplings.
Its least compelling feature is the fact that it has
over 100 new parameters and yet the current bound on the Higgs mass
requires fine-tuning of parameters for successful electroweak symmetry
breaking.  Models that improve this situation exist, and thus weakly
coupled superpartners at the weak scale remains an interesting option.
However, as we close in on the parameter space of the minimal supersymmetric
standard model, it is of great interest to find alternative weakly coupled 
theories of electroweak symmetry breaking.

A new class of models, called ``Little Higgs'' theories [3-8], 
produce a light Higgs boson with weakly coupled physics up to
10s of TeV.  As in composite Higgs models \cite{GK}, the Higgs
is a pseudo-Nambu-Goldstone boson (PNGB) and is massless at tree-level.
Its mass is protected by a global symmetry which is spontaneously broken.
The symmetry is explicitly broken by weakly coupled operators in 
the theory which become the normal SM couplings below 1 TeV.  The
Little Higgs trick is that no one operator alone explicitly breaks the
global symmetry protecting the Higgs mass, and therefore
no quadratically divergent contribution exists at one loop.  The purpose
of this article is to present a new model of this type with the simplest
gauge group to date.

In the next section, we show that simply extending the electroweak gauge
group to $SU(3)\times U(1)$ {\it automatically}
removes one-loop quadratic divergences from the gauge and top loops.  We'll
show that if two scalar triplets have vacuum expectation values (VEVs) in 
the same direction, breaking the gauge group to $SU(2)\times U(1)$, 
there exists an $SU(2)$ doublet whose mass is protected from
quadratic divergences at one loop.
Adding the minimum (weakly coupled) top sector to generate
a Yukawa coupling automatically protects the Higgs mass at one loop.
The key is that by having both triplets get VEVs, both produce a set of
NGBs (including doublets), either of which could have been eaten by the 
massive gauge fields.  It is the fact that no one operator in the tree-level 
Lagrangian contains both triplets which protects the doublet mass.
The symmetry protecting the Higgs mass in this model is an approximate
$[SU(3)/SU(2)]^2$.  We discuss the non-linear sigma model version of this
theory and show how to get there from the linear sigma model.

In Section \ref{su4} we present a complete model,
including a quartic Higgs coupling, 
which has no quadratic divergences at one loop.  Extending
the gauge group to $SU(4)\times U(1)$ allows one to 
generate a quartic coupling through a vacuum misalignment
mechanism.  The non-linear sigma model is $[SU(4)/SU(3)]^4$.  The result
is a two Higgs doublet model with heavy $SU(4)/SU(2)$ gauge bosons and new
scalars at a TeV.

In Section 4 we discuss the spectrum of the $SU(4)$ model and indicate
some of the phenomenological constraints [10-14].
Because our model has no more gauge couplings than the SM, all couplings
and masses in the gauge sector are a function of the breaking scale $f$
and measured gauge couplings.  We also comment on flavor issues in these 
models.

In the last section we discuss
%our model in relation to existing Little Higgs models and
possibilities for future directions.

\section{Little Higgs from a simple gauge group}
\label{su3}

In this section we describe a very simple extension of the
SM which keeps the Higgs naturally light by employing the
little Higgs mechanism. The model contains new particles and
couplings which cancel the quadratic divergences from the
top quark loop and from the SM gauge interactions. This is
accomplished by enlarging the $SU(2)$ weak gauge
interactions to $SU(3)$. The cancellation of the divergences from
the Higgs self-coupling is more difficult and is described in
Section 4.

Our mechanism for eliminating the
one-loop divergence from gauge interactions (Figure 1.b) can
be understood as follows:
an $SU(3)$ gauge group is spontaneously broken near 1 TeV by a
vacuum expectation value (VEV) of two $SU(3)$ triplet scalars.
The triplet expectation
values are aligned so that both vevs leave
the same $SU(2)$ unbroken. Ignoring their coupling trough the
gauge interactions each scalar breaks a global $SU(3)\rightarrow SU(2)$,
each yielding 5 Nambu-Goldstone bosons (NGBs).
With the $SU(3)$ gauge interactions turned on, the diagonal linear
combination of NGBs is eaten, but the
orthogonal linear combination remains massless at tree level.
The masslessness of these modes is easily understood by noticing
that in absence of a direct coupling between the two scalar triplets
each of them ``thinks'' that it is the only field which breaks the
gauge group and therefore contains exact NGBs. 
Quantum corrections from loops involving the gauge bosons generate
couplings between the two triplets and therefore a mass for the
pseudo-Nambu-Goldstone bosons (PNGB). However, there is no
quadratically divergent one-loop diagram involving both scalar triplets.
Finite and log-divergent diagrams contribute small scalar masses
of order $g/4\pi \ f \sim 100$ GeV.

More concretely, consider an $SU(3)$ gauge theory with two scalar
fields transforming as (complex) triplets 
${\bf \Phi}_i,\ i=1,2,$ of the gauge group with a potential
\beq
\frac{\lambda^2}2 ({\bf \Phi}_1^\dagger {\bf \Phi}_1 -f^2)^2\ + 
\ \frac{\lambda^2}2 ({\bf \Phi}_2^\dagger {\bf \Phi}_2 -f^2)^2
\label{eq:phipot}
\eeq
which generates VEVs for the ${\bf \Phi}$\,s.
For simplicity, we assume equal couplings and VEVs for both triplets.
This defines a linear sigma model with two global $SU(3)$
symmetries acting on the two triplets.
The spontaneous breaking $[SU(3)]^2 \rightarrow [SU(2)]^2$
yields five NGBs from each scalar. We now weakly gauge
an $SU(3)$ such that both scalars are triplets under the gauge symmetry.
This explicitly breaks the $[SU(3)]^2$ global symmetry to
diagonal $SU(3)$.
After spontaneous symmetry breaking the five ``diagonal''
NGBs are eaten by the Higgs mechanism and
the five orthogonal linear combinations are PNGBs.
The symmetry which they correspond to -- ``axial'' $SU(3)$ -- is
explicitly broken by the gauge couplings. But since the tree level
scalar potential respects both $SU(3)$'s the PNGBs
remain massless at tree level.

We parametrize the scalars as
\bea
{\bf \Phi}_1 = e^{i \Theta_1/f}
\left( \begin{array}{c} 0  \\ 0 \\ \!f\!+\!\rho_1\! \end{array} \right)
= e^{i \Theta_{eaten}/f}\ e^{i \Theta/f}
\left( \begin{array}{c} 0  \\ 0 \\ \!f\!+\!\rho_1\! \end{array} \right) 
\nonumber \\
{\bf \Phi}_2=e^{i \Theta_2/f}
\left( \begin{array}{c} 0  \\ 0 \\ \!f\!+\!\rho_2\! \end{array} \right)
= e^{i {\Theta_{eaten}/ f}} e^{-i \Theta/f}
\left( \begin{array}{c} 0  \\ 0 \\ \!f\!+\!\rho_2\! \end{array} \right)
\eea
where the first parametrization is the most obvious, but the
second is more convenient because it separates the eaten modes
$\Theta_{eaten}$ from the PNGBs $\Theta$. Note
that $\Theta_{eaten}$ shifts under diagonal (vector) $SU(3)$
transformations whereas $\Theta$ shifts under ``axial'' $SU(3)$.
The $\rho_i$ are radial modes which obtain masses
$m_\rho=\lambda f$ from the potential. Since we are
interested in the physics of the PNGBs we
will suppress the eaten fields $\Theta_{eaten}$.
Furthermore we will take the limit $\lambda \rightarrow 4 \pi$
in which the radial modes decouple and our linear sigma
model turns into the corresponding non-linear sigma model.
The non-linear sigma model obtained by integrating
out the $\rho_i$ includes non-renormalizable interactions
which become strongly coupled at
$\Lambda = 4\pi f$; at this scale the non-linear sigma
model description breaks down. For most of this paper we will
use the non-linear sigma model description because it is
more general. It focuses on the physics of the PNGBs.
Details of the UV theory which lead to the $[SU(3)]^2$ symmetry breaking
are encoded in higher dimensional operators and decouple
from the relevant physics.
The non-linear sigma model field are then parameterized as \cite{CCWZ}
\bea
\Phi_1= e^{i \Theta/f}
\left( \begin{array}{l}
0  \\ 0 \\ f \end{array} \right)\ , \quad 
\Phi_2= e^{-i \Theta/f}
\left( \begin{array}{l}
0  \\ 0 \\ f\end{array} \right)
\eea
where we suppressed the eaten fields and removed the bold type
to distinguish non-linear from linear sigma model fields.

Expanded out in components, the five PNGBs in
$\Theta$ are
\bea
\Theta= \Theta^a T^a =  
{1 \over \sqrt{2}}
\left( \begin{array}{cc} 
\!\!\begin{array}{ll} 0 & 0 \\ 0 & 0 \end{array} 
& \!\!h \\ h^\dagger & \!\!0 \end{array} \right)
+{\eta \over 4} 
\left( \begin{array}{rrr} 1&0 &0 \\0 &1&0\\ 0& 0&\!\!\!-2 \end{array}
\right)  \ .
\eea
The $T^a$ are the usual $SU(3)$ generators, $a$ runs
from $4 \dots 8$, and normalizations were chosen such that $h$ and $\eta$ 
have canonical kinetic terms. Under the unbroken $SU(2)$ gauge
symmetry $h$ transforms like the SM Higgs, i.e. it is a complex doublet,
and $\eta$ is a neutral scalar. 

Lets return to the linear sigma model.
Since the gauge interactions explicitly break the axial $SU(3)$ symmetry
which protects the PNGBs we expect that quantum
corrections from gauge interactions will generate a mass for them.
In the linear sigma model the $SU(3)$ gauge interactions are
\beq
 \left|(\partial_\mu+i g A_\mu){\bf \Phi}_1\right|^2+
 \left|(\partial_\mu+i g A_\mu){\bf \Phi}_2\right|^2 \ .
\label{eq:lineargauge}
\eeq
\begin{figure}[htb]
\vskip 0.0truein
\centerline{\epsfysize=1.4in
{\epsffile{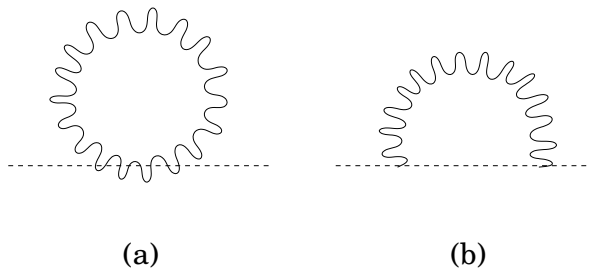}}}
\vskip 0.0truein
\caption[]{\it $SU(3)$ gauge boson loop corrections to the
${\bf \Phi}$ mass. }
\label{fig:loop2}
\end{figure}
At one loop, there are two quadratically divergent diagrams which contribute
to scalar masses (Figure 2). For the following arguments we
choose $\partial_\mu A^\mu=0$ gauge. This is convenient because
diagrams involving the trilinear gauge couplings cannot contribute
to the scalar potential (no derivatives) in this gauge. Thus the second
diagram vanishes.

Diagrams contributing to the
scalar potential are shown in Figure 3.
The first diagram is quadratically divergent, the
second one is log-divergent, and diagrams with even more ${\bf \Phi}$
insertions would be finite.
\begin{figure}[htb]
\vskip 0.0truein
\centerline{\epsfysize=1.6in
{\epsffile{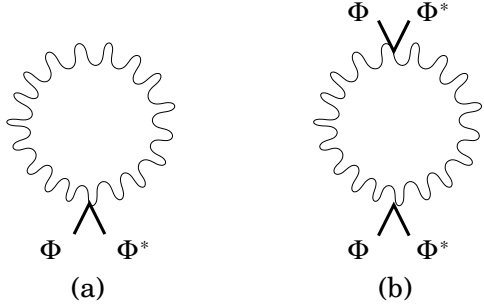}}}
\vskip 0.0truein
\caption[]{\it Gauge boson contributions to the ${\bf \Phi}$
potential in $\partial_\mu A^\mu =0$ gauge. }
\label{fig:loop3}
\end{figure}
The first diagram gives
\beq
\triangle {\cal L} \sim -{g^2\, \Lambda^2 \over 16 \pi^2}
( {\bf \Phi}_1^\dagger {\bf \Phi}_1 + {\bf \Phi}_2^\dagger {\bf \Phi}_2) \ ,
\eeq
which preserves both $SU(3)$ symmetries and renormalizes
the ${\bf \Phi}_i$ potentials, eq.~(\ref{eq:phipot}).
Again, taking $\Lambda=4 \pi f$ 
this contribution is not larger than the tree level terms
already present in the potential and does not destabilize the
desired vacuum.
Thus at the level of quadratically divergent diagrams, the 
PNGBs remain massless.

Log-divergent and finite diagrams do contribute to
PNGB masses. For example, the second diagram in Figure 3
generates
\bea
\triangle {\cal L} \sim {g^4 \over 16 \pi^2 }
|{\bf \Phi}_2^\dagger {\bf \Phi}_1|^2\,
{\rm log}(\Lambda^2/f^2) 
\ \sim \ -{f^2 \over 16 \pi^2 } \ h^\dagger h \ .
\eea
This operator explicitly violates axial $SU(3)$
and contains a mass for the PNGBs. 
However,
the mass is of order $f/4\pi$ which is sufficiently small. 
Note that the mass-squared generated is positive, which means
that this term stabilizes the vacuum with aligned expectation
values for ${\bf \Phi}_1$ and ${\bf \Phi}_2$. The top
quark loop which we discuss in the next section generates
a negative mass-squared and thereby triggers electroweak
symmetry breaking.

This analysis could have also been performed in the
non-linear sigma model. The expression
for the gauge couplings is identical to eq.~(\ref{eq:lineargauge})
except that the ${\bf \Phi}_i$ are replaced by $\Phi_i$.
Loops like Figure 3.a generate
$|\Phi_i^\dagger  \Phi_i|=f^2$, a quadratically divergent
contribution to the cosmological constant but no mass.
Finite and log-divergent diagrams generate
$|\Phi_2^\dagger \Phi_1|^2$ which does contain a
Higgs mass of order 100 GeV if $f\sim$ 1 TeV. 

The absence of quadratic divergences can also be understood
by noting that the diagram in Figure 3.a  only involves
one of the non-linear sigma model fields.
Thus it is identical to the corresponding diagram in a theory
with only one $\Phi$. But in this theory $\Theta$ would be
eaten by the Higgs mechanism, thus it cannot
have any non-derivative couplings in the Lagrangian.
Therefore, only diagrams which involve both $\Phi$'s (Figure 3.b)
can contribute
to the Higgs mass but they necessarily involve more internal
propagators and are not quadratically divergent.

\subsection{Top Yukawa coupling}

We now show that it is straightforward to add fermions and a top Yukawa
coupling which does not upset the radiative stability of
the Higgs mass. This can be done for both the linear and non-linear
sigma models but we will only present the analysis of the non-linear
model.

Since $SU(2)$-weak is embedded into an $SU(3)$ gauge group,
the top-bottom $SU(2)$-doublet is enlarged to a
triplet $Q^T=(t,b,\chi)$. A mass of order $f$ for the extra fermion
$\chi$ in the triplet and Yukawa couplings for the top quark
are generated from couplings to the $\Phi$'s
\bea
{\cal L}_{top} =
\lambda_1 \chi_{1}^c \Phi_1^\dagger Q +
\lambda_2 \chi_{2}^c \Phi_2^\dagger Q \ ,
\label{eq:topyuk}
\eea
where $\chi_{i}^c$ are Weyl fermions with the quantum numbers
of the $SU(2)$-singlet component of the top quark.%
\footnote{Note that despite superficial similarities between
eq.~(\ref{eq:topyuk}) and the top-color see-saw \cite{topseesaw}
there is an important difference. Unlike the top-$\chi$ Higgs couplings
in \cite{topseesaw}, the couplings in eq.~(\ref{eq:topyuk})
``collectively'' break an $SU(3)$ symmetry protecting the Higgs mass.
This is necessary for the cancelation of quadratic divergences
to occur, and this is why we obtain a top Yukawa coupling which does
not require fine tuning of the Higgs mass.}  
The Yukawa couplings $\lambda_i$ can be chosen real by redefining
the phases of the $\chi^c$. Expanding to first order in the Higgs $h$
\bea
{\cal L}_{top} &=&
 f (\lambda_1 \chi_{1}^c + \lambda_2 \chi_{2}^c) \chi
+ {i \over \sqrt{2}}
(\lambda_1 \chi_{1}^c - \lambda_2 \chi_{2}^c) h
\left( \begin{array}{l} t  \\ b \end{array} \right) + \cdots \\
&=& m_\chi \ \chi^c \chi
- i \lambda_t \ t^c\, h
\left( \begin{array}{l} t  \\ b \end{array} \right)
+ \cdots \ ,
\eea
where in the last step we diagonalized the mass matrix by finding the
heavy and light linear combinations of the $\chi^c$. We find
a $\chi$ mass 
$m_\chi = \sqrt{\lambda_1^2+\lambda_2^2}\, f$, and 
a top Yukawa coupling, $\lambda_t=\sqrt{2}\, \lambda_1 \lambda_2 /
\sqrt{\lambda_1^2+\lambda_2^2}$.
To obtain a sufficiently large top mass
both couplings $\lambda_i$ must be of order one.

The absence of quadratic divergences to the Higgs mass at
one loop from these interactions is again most easily understood
by examining Feynman diagrams with external $\Phi$'s. The
diagram in Figure 4.a is quadratically divergent but
preserves both $SU(3)$ symmetries. Thus it does not contribute to
PNGB masses. The diagram 4.b does contribute to
the Higgs mass, but it is only log divergent, contributing to the
operator
$\lambda_1^2 \lambda_2^2/16\pi^2 |\Phi_2^\dagger \Phi_1|^2$
which contains a Higgs mass of order $f \lambda /4\pi$.
\begin{figure}[htb]
\vskip 0.2truein
\centerline{\epsfysize=1.8in
{\epsffile{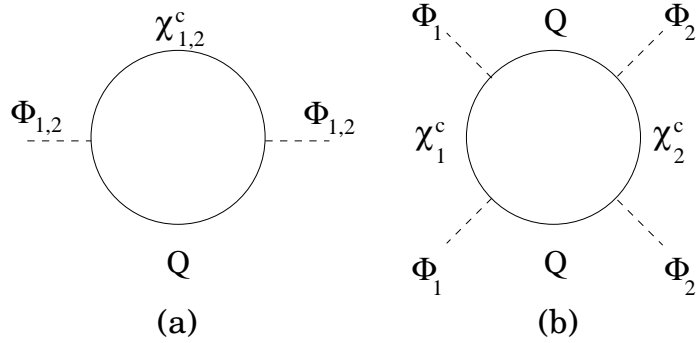}}}
\vskip 0.2truein
\caption[]{\it The top loop contribution to the Higgs mass in
sigma model formalism. }
\label{fig:topphi2}
\end{figure}

Alternatively, one can perform this computation after
expanding the $\Phi$'s. In ``component form'' the vanishing
of the quadratic divergence involves ``miraculous'' cancellations
between loops of top quarks and loops of $\chi$'s. We demonstrate this
calculation for the simplifying choice of
$\lambda_1=\lambda_2\equiv\lambda_t$.
Expanded out to the relevant order, the Lagrangian contains
\beq
\lambda_t 
(\sqrt{2}f  - \frac1{\sqrt{2}f} h^\dagger h ) \chi^c \chi
+ \lambda_t t^c h
\left( \begin{array}{l} t  \\ b \end{array} \right)\ .
\eeq
In addition to the usual top loop there is also a quadratically divergent
$\chi$ loop (Figure 5).
In the diagram the $\lambda_t \sqrt{2}f$ mass insertion on the
$\chi$ line combines with the $-\lambda_t/ \sqrt{2}f$ from the vertex
to exactly cancel the quadratic divergence from the top loop.
\begin{figure}[htb]
\vskip 0.2truein
\centerline{\epsfysize=1.8in
{\epsffile{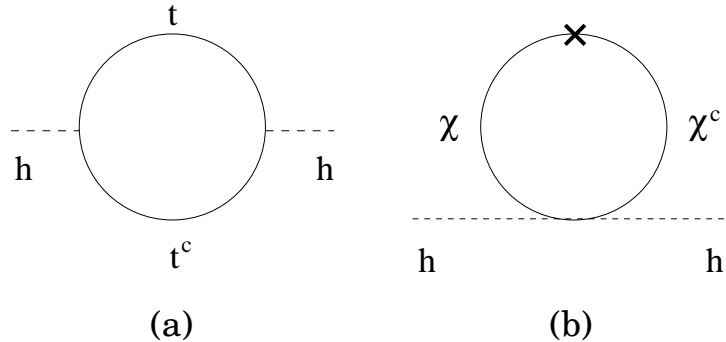}}}
\vskip 0.2truein
\caption[]{\it The canceling top and $\chi$-loops in component form. }
\label{fig:chiloop}
\end{figure}
Note that although the cancellation involves a higher dimensional operator,
its coefficient is related to the Yukawa coupling by the
non-linearly realized axial $SU(3)$ symmetry.

\subsection{$[SU(3)/SU(2)]^2$ symmetry}

Here we give an elegant way of understanding the absence of
quadratic divergences in our theory which relies on the structure
of the explicit breaking of the $[SU(3)]^2$ symmetry acting on the
$\Phi$ fields. The $[SU(3)]^2$ symmetry not only protects both sets
of NGBs ($\Theta_{eaten}$ and $\Theta$) from obtaining
a mass, it also forbids any non-derivative couplings of the
NGBs. Therefore, in order to generate gauge- and Yukawa couplings
for the Higgs, these symmetries must be explicitly broken. The
trick is to do this breaking ``non-locally in theory space''
\cite{lh} or ``collectively''. By collective breaking
we mean that no single coupling in the Lagrangian breaks the symmetry
by itself. Always at least two couplings are required
to break any of the two $SU(3)$'s.

To see this explicitly, consider the gauge interactions
\beq
 \left|(\partial_\mu+i g_1 A_\mu)\Phi_1\right|^2+
 \left|(\partial_\mu+i g_2 A_\mu)\Phi_2\right|^2 \ ,
\eeq
where for clarity we have labeled the gauge couplings of
$\Phi_1$ and $\Phi_2$ differently. Of course, gauge invariance requires
them to have the same value, but as spurions they break
different symmetries, and it is useful to keep
track of them independently.

The key is to note that if either of the two couplings $g_1$ and $g_2$
is set to zero the Lagrangian has an exact $[SU(3)]^2$ symmetry, and
therefore two sets of exact NGBs ($\Theta_{eaten}$ and $\Theta$)
result from the spontaneous symmetry breaking.
Only with both $g$'s non-vanishing are the symmetries
explicitly broken to the diagonal gauged $SU(3)$.
For $g_2=0$ we have the two independent
symmetries
\bea
\Phi_1\rightarrow U_1 \Phi_1 , \quad 
A_\mu \rightarrow U_1 A_\mu U_1^\dagger , \quad
\Phi_2\rightarrow U_2 \Phi_2 \ ,
\eea
while for $g_1=0$ we have the symmetries
\bea
\Phi_1\rightarrow U_1 \Phi_1 , \quad 
A_\mu \rightarrow U_2 A_\mu U_2^\dagger , \quad
\Phi_2\rightarrow U_2 \Phi_2 \ .
\eea
Thus when either of the $g_i$ is set to zero $\Theta$ is an
exact NGB. Any loop correction
to its mass must be proportional to both $g_1$ and
$g_2$. But, there are no quadratically divergent one loop diagrams
involving both $g_1$ and $g_2$.

The argument for the Yukawa couplings is very similar. In absence
of either $\lambda_1$ or $\lambda_2$ the Yukawa couplings
in eq.~(\ref{eq:topyuk}) preserve two $SU(3)$ symmetries.
For example, when $\lambda_2=0$ we have
\bea
\Phi_1\rightarrow U_1 \Phi_1 , \quad 
Q \rightarrow U_1 Q , \quad
\Phi_2\rightarrow U_2 \Phi_2 \ ,
\eea
where $Q$ denotes the quark triplet. Thus
any $\Theta$ mass which is generated from Yukawa loops must be
proportional to both $\lambda_1$ and $\lambda_2$. In fact, it must be
proportional to $|\lambda_1|^2|\lambda_2|^2$. This follows from the
two spurious $U(1)_i$ symmetries under which
only $\lambda_i$ and $\chi_{i}^c$ are charged such that
$\lambda_i \chi_{i}^c$ are neutral. Any operator which
is generated by loops must also respect these spurious symmetries,
and if it doesn't contain $\chi$\,s it can only depend on the
invariant combinations $|\lambda_1|^2$ and $|\lambda_2|^2$.
Again, there is no quadratically divergent one-loop diagram proportional
to $|\lambda_1|^2|\lambda_2|^2$.

\subsection{Lets make the Standard Model}

Three ingredients are still missing to turn the above model
into a fully realistic extension of the Standard Model. We
need {\it i.} hypercharge and color, {\it ii.} Yukawa couplings
for the other fermions, and {\it iii.} a quartic self coupling for
the Higgs in order to stabilize its VEV.

\noindent {\it i. Hypercharge and color:} Adding color is trivial.
To one loop the only relevance of color for the
Higgs mass is a color factor of three
in the top quark loop. Adding hypercharge is also straightforward.
We introduce it by gauging a $U(1)_X$ symmetry which
commutes with the $SU(3)$ and under which the
$\Phi_i$ have charge $-1/3$. In the non-linear sigma model $U(1)_X$ is
non-linearly realized but the linear combination corresponding to
hypercharge
\bea
Y=\frac{1}{\sqrt{3}}\, T_8\, +\, \frac13 X
\eea
is unbroken.
Here $T_8=\frac{1}{\sqrt{3}}\,diag(-\frac12,-\frac12,1)$
is one of the broken $SU(3)$ generators. The $U(1)_X$ charges of the fermions
are uniquely determined from their hypercharges. For example,
$\chi_i^c$ has charge $-\frac23$ and $Q$ has charge $\frac13$.

Gauging $U(1)_X$ does not introduce new 
quadratic divergences for the Higgs mass. This is because
$U(1)_X$ commutes with all the spurious global $SU(3)$ symmetries
which we used to argue for the absence of divergences. Thus
these arguments go through unchanged.

\noindent{\it ii. Yukawa couplings for light fermions:} Since the Yukawa
couplings of the other fermions are small, no care needs to
be taken in their coupling to the Higgs in order to avoid
large Higgs mass corrections. One possibility is to enlarge all
left-handed fermion doublets into triplets and then write
Yukawa couplings for up-type quarks and neutrinos just as
we did for the top quark. For the bottom quark
we can write
\bea
{\lambda_b\over \Lambda}\ b^c
\epsilon_{ijk} \Phi_1^i \Phi_2^j\, Q^k + h.c. \ ,
\eea
and analogous
terms for all down-type quarks and charged leptons. 
We postpone a more detailed discussion of flavor until Section 4.

{\it iii. Higgs quartic coupling:} 
One possibility for the quartic coupling is to ignore the numerically
not very significant one-loop divergence from the Higgs loop and
add a quartic coupling without canceling its divergence. A
more satisfying model in which this divergence is also canceled
requires more work.
It is easy to appreciate the difficulty
by looking at possible terms which one might add to the Lagrangian to
generate the quartic coupling. This would-be Higgs potential is an
arbitrary gauge invariant polynomial in $\Phi_1$ and $\Phi_2$.
The only non-trivial gauge invariant contraction at our disposal is
$\Phi_1^\dagger \Phi_2$.
The others either vanish ($\epsilon_{ijk} \Phi_1^i \Phi_1^j \Phi_2^k=0$)
or are constant ($\Phi_i^\dagger \Phi_i=f^2$). Thus the
potential is a function of
\bea
\Phi_1^\dagger \Phi_2 = f^2 +i f \eta - h^\dagger h -\frac12 \eta^2
+ \cdots + \frac1{6 f^2} (h^\dagger h)^2 + \cdots
\label{eq:phi1phi2}
\eea
and it's hermitian conjugate. Focusing on the $h$-dependence, we see that
the quartic coupling always comes accompanied by a mass term when
expanding out a term like $(\Phi_1^\dagger \Phi_2)^n$. Setting the
coefficient of the quartic to one, the mass is of order $f$ which is
much too large. Note that if the constant term in the expansion
of $\Phi_1^\dagger \Phi_2$ were not there, one could generate
a quartic by simply squaring eq.~(\ref{eq:phi1phi2}). This observation
will be the key to constructing a quartic in Section 4.
Of course, it is possible to fine tune the mass
term away. For example the potential $|\Phi_1^\dagger \Phi_2-f^2|^2$
does not contain a mass term. But this is no better than
the tuning of the Higgs mass in the Standard Model as the
relative size of coefficients in $|\Phi_1^\dagger \Phi_2-f^2|^2$
is not stable under quantum corrections.

\section{A Complete Model: SU(4)}
\label{su4}

Now we present a complete model of electroweak symmetry breaking with
no fine-tuning required to separate the Higgs mass from the cutoff.
The main distinction from the previous model is the extension of the
gauge group (and therefore the approximate global symmetries) from 
$SU(3)$ to $SU(4)$.  The expansion of the group allows one to use a new
mechanism to generate a quartic coupling.

Take $\Phi$ and $\Psi$ as fields in the non-linear sigma model
$[SU(4)/SU(3)]^2$ with the diagonal $SU(4)$ gauged.  The important distinction
from the previous model is that the $SU(4)$-breaking is {\it not} aligned,

\bea
\Phi= e^{i \varphi /f}
\left( \begin{array}{l}
0  \\ 0 \\ f \\ 0 \end{array} \right) \\
\Psi=e^{i {\psi /f}} 
\left( \begin{array}{l}
0  \\ 0 \\ 0 \\ f \end{array} \right)
\eea
and only the gauged $SU(2)$ is linearly realized.  Note that the product
$\Phi^{\dagger} \Psi$ contains no constant term!
This is exactly the success we were attempting to achieve in the $SU(3)$ model.
Raised to the appropriate power, it could potentially contain a term
quartic in Higgses without a quadratic term.

This (mis-)alignment is stabilized when one adds the interaction 
$|\Phi^{\dagger} \Psi |^2$ to the potential with a positive 
coefficient.  As we will see, this interaction becomes a positive 
squared mass for a charged scalar field -- the only uneaten (complex) 
scalar in this example.  The two $SU(2)$ doublets which live in 
$\Phi$ and $\Psi$ remain exact NGBs (they are in fact eaten) 
and thus this term does not induce a quartic term.

To reproduce the successes of the $SU(3)$ model (and produce uneaten Higgs
doublets), we break the gauged $SU(4)\rightarrow SU(2)$ twice.
Our model has four sets of sigma model fields, ($\Phi_i, \Psi_i,\ i=1,2$).
Each contains one complex $SU(2)$ doublet.  Of the four doublets, two
are eaten by the heavy $SU(4)$ gauge bosons, leaving a two-Higgs-doublet
model.  

The complete counting goes as follows:  the $[SU(4)/SU(3)]^4$ 
represents $(15 - 8) \times 4 = 28$ real components,
12 of which are eaten when the $SU(4)$ gauge group is broken to $SU(2)$.  
The remaining 16 consist of two complex doublets $h_u$ and $h_d$, 
three complex $SU(2)$ singlets $\sigma_1$, $\sigma_2$
and $\sigma_3$, and two real scalars $\eta_u$ and $\eta_d$.
One possible parameterization is as follows: 
\bea
\Phi_1=e^{+i {\cal H}/f} e^{+i \Sigma_1 /f} e^{+i \Sigma_2/f} e^{+i \Sigma_3/f}
        e^{+ i \eta_u / f} 
        \left( \begin{array}{l} 0  \\ 0 \\ f \\0 \end{array} \right) \\
\Phi_2=e^{- i {\cal H}/f} e^{+i \Sigma_1 /f} e^{-i \Sigma_2/f} e^{- i \Sigma_3/f}
        e^{ - i \eta_u / f}
        \left( \begin{array}{l} 0  \\ 0 \\ f \\0 \end{array} \right) \\
\Psi_1=e^{+i {\cal H}/f} e^{- i \Sigma_1 /f} e^{+i \Sigma_2/f} e^{- i \Sigma_3/f}
        e^{ + i \eta_d / f}
        \left( \begin{array}{l} 0  \\ 0 \\ 0 \\f \end{array} \right) \\
\Psi_2=e^{- i {\cal H}/f} e^{- i \Sigma_1 /f} e^{-i \Sigma_2/f} e^{+i \Sigma_3/f}
        e^{ - i \eta_d / f}
        \left( \begin{array}{l} 0  \\ 0 \\ 0 \\f \end{array} \right)
\eea
with
\bea
  {\cal H} =
    {1 \over \sqrt{2}}
    \left( \begin{array}{cccc} 
      \!\!\begin{array}{ll} 0 & 0 \\ 0 & 0 \end{array} 
      & \!\!h_u & \!\! h_d \\ 
      h_u^\dagger & \!\!0 & \!\! 0 \\
      h_d^\dagger & \!\! 0 & \!\!0 \\ 
    \end{array} \right) &&
  \Sigma_1 = 
    {1 \over 2}
    \left( \begin{array}{cccc}
      0 & 0 & 0 & 0 \\ 0 & 0 & 0 & 0 \\ 0 & 0 & 0 & \sigma_1 \\
      0 & 0 & \sigma_1^\dagger & 0 \\
    \end{array} \right)\nonumber\\
        \nonumber\\
 \Sigma_2 =
    {1 \over 2}
    \left( \begin{array}{cccc}
      0 & 0 & 0 & 0 \\ 0 & 0 & 0 & 0 \\ 0 & 0 & 0 & \sigma_2 \\
      0 & 0 & \sigma_2^\dagger & 0 \\
    \end{array} \right) &&
  \Sigma_3 = 
    {1 \over 2}
    \left( \begin{array}{cccc}
      0 & 0 & 0 & 0 \\ 0 & 0 & 0 & 0 \\ 0 & 0 & 0 & \sigma_3 \\
      0 & 0 & \sigma_3^\dagger & 0 \\
    \end{array} \right) \\
        \nonumber\\
  {\bf\rm \eta}_u = 
    {\eta_u \over 6}
    \left( \begin{array}{rrrr} 1&0 &0 &0 \\0 &1&0&0\\0 &0&\!\!\! -3&0\\ 
      0& 0& 0&\!\!\!1 
    \end{array} \right) &&
  {\cal \eta}_d =
    {\eta_d \over 6}
    \left( \begin{array}{rrrr} 1&0 &0 &0 \\0 &1&0&0\\0 &0&\!\!\! 1&0\\ 
      0& 0& 0&\!\!\! -3 
    \end{array} \right) \nonumber\ .
\eea
A quartic term for this model will come from the four couplings 
\bea
\kappa_{11} |\Phi_1^{\dagger} \Psi_1 |^2 
+ \kappa_{22} |\Phi_2^{\dagger} \Psi_2 |^2 
+ \kappa_{12} |\Phi_1^{\dagger} \Psi_2 |^2 
+ \kappa_{21} |\Phi_2^{\dagger} \Psi_1 |^2 \ .
\eea   
To leading order in each field, these operators produce the potential:
\bea
\kappa_{11} f^2 | \sigma_1 + \sigma_3|^2 
&+& \kappa_{22} f^2 | \sigma_1 - \sigma_3|^2 \nonumber\\
+ \kappa_{12} f^2 | \sigma_1 + \sigma_2 - i h_u^{\dagger} h_d/f|^2 \nonumber
&+& \kappa_{21} f^2 | \sigma_1 - \sigma_2 - i h_u^{\dagger} h_d/f|^2 \ .
\eea
Any one coupling does not produce a potential for the Higgses.  
In fact, removing any of the couplings above removes the quartic Higgs term.  
To see this, note that there are other parameterizations in which the 
$h_u^{\dagger} h_d$ appear in different operators.

These couplings generate masses of order $\sqrt{\kappa} f$
for the complex scalars $\sigma_i$ as well as trilinear
terms marrying these scalars to $h_u^\dagger h_d$.
Integrating out the singlets produces a quartic coupling:
\beq
\lambda = \frac{4}{\kappa_{11}^{-1} 
        + \kappa_{22}^{-1} + \kappa_{12}^{-1} + \kappa_{21}^{-1}}
\eeq
which vanishes as any one $\kappa_{ij}\rightarrow 0$, thus all four 
terms are required to produce a tree-level quartic term.

From symmetry arguments we see why this works.  The non-linearly 
realized global symmetry of the model is approximately 
$[SU(4)/SU(3)]^4$ and contains, among other things, four
Higgs-like doublets which are NGBs.  Two of the doublets are 
eaten.  A single operator, {\it e.g.}, $|\Phi_1^\dagger \Psi_1|^2$, 
explicitly breaks the symmetry down to $[SU(4)/SU(3)]^2 
\times [SU(4)/SU(2)]$, but this symmetry {\it also} contains four
doublet NGBs.  Adding $|\Phi_2^\dagger \Psi_2|^2$ leaves 
$[SU(4)/SU(2)]^2$ again producing four doublets.  The existence 
of three of the four operators breaks enough symmetry to allow 
a quartic term, but only a small one is induced at loop level.

This structure also suppresses one-loop contributions to the Higgs 
mass.  The symmetry arguments above show that more than one spurion 
is required to generate a Higgs potential, and therefore a Higgs mass.  
Thus, at one loop there are no quadratically divergent diagrams.

The gauge and top loops work similarly to those in the $SU(3)$ model.  
The gauge loop is canceled for each Higgs doublet by the massive 
gauge bosons and the calculation goes through as in Section \ref{su3}.  
Gauging an additional $U(1)$ symmetry results in the existence of 
hypercharge at the weak scale and does not contribute to the Higgs 
potential.  The top Yukawa coupling could, for example, come from:
\bea
{\cal L}_{top} = 
(\lambda_1 \chi_1^c \Phi_1^\dagger + \lambda_2 \chi_2^c \Phi_2^\dagger
  +\lambda_3 \chi_3^c \Psi_1^\dagger)\  Q
\eea
where $Q^T=(t, b, \chi_1, \chi_2)$.
The $\chi_3^c$ field is there to cancel the hypercharge anomaly 
and to marry and give a mass to $\chi_2$.  The $\lambda_3$ term 
does not play a role in generating a top Yukawa coupling and thus 
the coupling can be taken to the cutoff ({\it i.e.}, 
$\lambda_3 \rightarrow 4 \pi$) thus decoupling these extra fields 
and making the physics the same as the $SU(3)$ case of the previous 
section. The bottom Yukawa coupling can be written as
\bea
{\cal L}_{bottom} = 
\lambda_b b^c \; \epsilon_{ijkl} \; \Phi_1^i \Psi_1^j \Psi_2^k Q^l \ .
\eea
This single coupling generates a quadratic divergence to the down-type 
Higgs mass of the form $|\Psi_1^\dagger \Psi_2|^2$.  For a small bottom 
Yukawa coupling ($\lambda_b \lsim 0.1$), the contribution remains 
at or below the weak scale.

\subsection{Electroweak Symmetry Breaking}
The purpose of the quartic term in the Higgs potential is to stabilize 
the Higgs VEV .  The $SU(4)$ model above has a quartic potential of the form
\beq
{\cal L}_{quartic} = - \lambda |h_u^\dagger h_d |^2
\eeq
where $h_u$ and $h_d$ are $SU(2)$ doublets with hypercharge $Y=-1/2$.  
Successful electroweak symmetry breaking requires the existence of a 
mass term of the form 
$B h_u^\dagger h_d$ where $B$ is of order $M_W^2$.
Such a term comes from operators
$B_{11} \Phi_1^\dagger \Psi_1$,
$B_{22} \Phi_2^\dagger \Psi_2$,
$B_{12} \Phi_1^\dagger \Psi_2$,
and $B_{21} \Phi_2^\dagger \Psi_1$,
and can be written with the quartics as
$\kappa_{11} | b_{11} + \Phi_1^\dagger \Psi_1|^2 +
\kappa_{22} | b_{22} + \Phi_2^\dagger \Psi_2|^2 + \dots$.
These new operators contain
linear terms for the uncharged scalar fields $\sigma_i$ causing them to 
obtain VEVs.  When the $\sigma$'s are shifted to their minima, a mass term
of the form $B h_u^\dagger h_d$ is produced with
\beq
B= {1\over 2} \lambda \sum_{ij} b_{ij} \ .
\eeq
Here -- for simplicity -- we have taken the $b_i$ to be real, though in
general their phases may have interesting implications for CP violation. 
These $B$ terms are themselves spurions
which explicitly break more symmetries than the quartic couplings.  To
see this, note that they are the only terms so far which produce a potential
for the $\eta$ fields.  Thus their size, which needs to be
of order $f^2/16 \pi^2$ is technically natural, but undetermined 
from dynamics in the effective theory below $\Lambda$.

In addition, operators of the form $|\Phi_1^\dagger \Phi_2|^2$ and 
$|\Psi_1^\dagger \Psi_2|^2$ are generated by two-loop quadratic-divergent and
one-loop log-divergent diagrams.  They produce the mass terms
\beq
{\cal L}_{mass} = m_2^2 |h_u |^2 + m_1^2 |h_d |^2 
\eeq
with a natural size of order $f^2/16\pi^2$.  We require $m_2^2 , m_1^2 > 0$
(or else the Higgs vev could run away to $\sim f$).  Electroweak symmetry
breaking occurs if 
\begin{equation}
B > \sqrt{m_2^2 m_1^2}\ .
\end{equation}
This Higgs potential is of the same form as the one in the 
$SU(6)/Sp(6)$ little Higgs model \cite{Low}, and we repeat
some of the phenomenology here. Formulas for the scalar masses in
general two Higgs doublet models are conveniently collected
in \cite{howie}. Minimizing the
potential under these conditions gives $\tan{\beta} = v_u/v_d \equiv
\langle h_u \rangle / \langle h_d \rangle = \sqrt{m_1^2/m_2^2}$ and
\beq
\frac{2 B}{\sin{2\beta}} = m_2^2 + m_1^2 + 2 \lambda v^2
\eeq
where $v = 174$ GeV is the electroweak symmetry breaking scale.

The masses of the two CP-even Higgs bosons are
\beq
M_{h^0}^2 , M_{H^0}^2 = 
                \frac{B}{\sin{2\beta}} 
                \pm \sqrt{\frac{B^2}{\sin^2{2\beta}} 
                        + \lambda v^2 
                        \left(\lambda v^2 - \frac{2 B}{\sin{2\beta}}\right)
                        \sin^2{2\beta} 
                } \ .
\eeq
The lightest CP-even Higgs boson is bounded from above by 
$M_{h^0}^2 \leq \lambda v^2$.  This bound is saturated for
$m_1^2 = m_2^2 \rightarrow \sin{2\beta}=1$.  The CP-odd and charged 
Higgses have masses
\bea
M_{A^0}^2 &=& \frac{2 B}{\sin{2\beta}} \\
M_{H^{\pm}}^2 &=& M_{A^0}^2 - \lambda v^2
\eea

\subsection{The Standard Model Embedding}
Now we can construct a complete standard model based on the $SU(4)$ theory.
Collecting the pieces together, the Lagrangian of the theory is
\beq
{\cal L} = {\cal L}_{kinetic} + {\cal L}_{quarks} + {\cal L}_{leptons} + {\cal L}_{Higgs}
\eeq
with
\beq
{\cal L}_{kinetic} = |{\cal D}_{\mu}  \Phi_i |^2 + 
     |{\cal D}_{\mu}  \Psi_i|^2 + \left[{\rm fermion\: and\:
    gauge\: kinetic\: terms}\right] \ .
\eeq
Here ${\cal D}_{\mu}= (\partial_{\mu} + i g_4 A_{\mu}^a T^a - i {g_X\over 4}
A_{\mu}^X)$ and the $\frac14$ in the coupling of $A_\mu^X$
represents the $U(1)_X$ charge of $\Phi$ and $\Psi$.
$A_{\mu}$ and $A^X_{\mu}$ are the $SU(4)$ and $U(1)_X$
gauge fields respectively.
The Yukawa couplings for quarks appear as
\beq
{\cal L}_{quarks} = (\lambda_{1}^u\, \chi_{u1}^{c}\, \Phi_1^\dagger  
        + \lambda_2^{u}\; \chi_{u2}^{c}\, \Phi_2^\dagger 
                + \lambda_3^{u}\, \chi_{u3}^{c}\, \Psi_1^\dagger) Q 
                +\: \lambda^d\; d^c \, \Phi_1 \Psi_1 \Psi_2 Q
\label{eq:quarkyuks}
\eeq
with $Q = (q, \chi_{u1}, \chi_{u2})^T$.
We have suppressed flavor and $SU(4)$ indices for
clarity.  The $\lambda$ couplings are $3\times 3$ matrices
in flavor space -- the combination
of the first three produces the standard Yukawa matrix
while $\lambda^d$ is simply the 
Yukawa matrix for down-type quarks.  Similarly, for the charged leptons:
\beq
{\cal L}_{leptons} = (\lambda_{1}^\nu\, \chi_{\nu1}^{c}\, \Phi_1^\dagger  
        + \lambda_2^{\nu}\; \chi_{\nu2}^{c}\, \Psi_1^\dagger ) Q 
         +\lambda^e\; e^c \, \Phi_1 \Psi_1 \Psi_2 L 
\eeq
where $L = (\ell, \chi_{\nu 1}, \chi_{\nu 2})^T$, we will discuss neutrino masses
in the next section. Finally, the tree level scalar potential is
\bea
{\cal L}_{\!scalar}\! = 
%\delta_u | \Phi_1^\dagger \Phi_2 |^2
%                + \delta_d | \Psi_1^\dagger \Psi_2 |^2 + \\
 \sum_{ij}   \kappa_{ij}\, | b_{ij} + \Phi_i^\dagger \Psi_j |^2 
 %   \kappa_2 | b_{2}\! +\! \Phi_2^\dagger \Psi_2 |^2 + 
 %   \kappa_3 | b_{3}\! +\! \Phi_1^\dagger \Psi_2 |^2 + 
 %   \kappa_4 | b_{4}\! +\! \Phi_2^\dagger \Psi_1 |^2
\eea
as discussed in the previous subsection.

Hypercharge is a linear combination of the $SU(4)$ generator 
\beq
T^{15} = \sqrt{2}\, diag(-1/4,-1/4,+1/4,+1/4)
\eeq
and the external $U(1)_X$.  Thus the
$X$ charges of the $SU(4)$-singlet fermions
are just their respective 
hypercharges while the $X$ charges of $SU(4)$ vectors are the hypercharges
of the $SU(2)$ doublets they contain plus $1/4$. 
Explicitly, $(\Phi_i,\Psi_i,L,Q)$ have $X$ charges
$(-1/4,-1/4,-1/4,+5/12)$.

\section{Spectrum and Constraints}
\label{pheno}

A complete analysis of the phenomenology is
beyond the scope of this paper but we would like to report on
our initial explorations in this direction. The results are encouraging:
we find significant constraints but there are large regions of 
parameter space which are in agreement with experiment
while at the same time solving the hierarchy problem.
Interestingly, the preferred region or parameter space will be
directly explored at the Tevatron and LHC.
More specifically, we will discuss {\it i.} precision electroweak
constraints {\it ii.} the spectrum and direct
searches {\it iii.} flavor physics.

\noindent {\it i. precision constraints:}
One of the most stringent constraints on models of new physics at
the TeV scale comes from isospin violating couplings
of the light fermions to the $W$ and $Z$ bosons.
One source of isospin violation is the different treatment of the
Yukawa couplings for up and down-type quarks in eq.~(\ref{eq:quarkyuks}).
Through the Higgs vev up-type quarks mix with the heavy $\chi$
fermions whereas down-type quarks don't. As we will now show,
this mixing prefers different scales $f_i$ for the
different $\Phi_i$.

For simplicity, we revert to our $SU(3)$ model where the same
mixing occurs. The up-type Yukawa couplings are
\beq
(\lambda_1 u^c_1 \Phi^\dagger _1 + 
\lambda_2 u^c_2 \Phi^\dagger _2 )\;
\left( \begin{array}{l}
u  \\ d \\ \chi\end{array} \right) \ ,
\eeq
where $\lambda_i$ are $3\times3$ matrices in flavor space. In order
to avoid large flavor changing effects (see FCNC discussion below)
we take $\lambda_2$ proportional to the unit matrix
and of order one whereas $\lambda_1$ is approximately equal to the
usual Yukawa couplings of the up-type quarks in the Standard Model.
Furthermore, we allow different scales $f_1$ and $f_2$ for the
non-linear sigma models. The heavy $SU(3)$
gauge bosons eat a linear combination of the NGBs
which resides mostly in the sigma model with the larger scale,
and the little Higgs lives mostly in the sigma model
with the smaller scale:
\bea
\Phi_1= e^{i\Theta  {f_2\over f_1} }
\left( \begin{array}{l}
0  \\ 0 \\ f_1 \end{array} \right) , \quad
\Phi_2= e^{-i \Theta {f_1\over f_2}}
\left( \begin{array}{l}
0  \\ 0 \\ f_2\end{array} \right) 
\eea
where 
\bea
\Theta = 
\left( \begin{array}{cc} 
\!\!\begin{array}{ll} 0 & 0 \\ 0 & 0 \end{array} 
& \!\!h \\ h^\dagger  & \!\!0 \end{array} \right)
/ f_{12}\quad  {\rm and}\quad  f_{12}^2 = f_1^2+f_2^2 \ .
\eea
Substituting the Higgs by its expectation value $h^T=(v,0)$ we obtain
the mass matrix
\bea
\left(\, u_1^c \  u^c_2 \right)
\left( \begin{array}{cc} 
\lambda_1 v \, \frac{f_2}{f_{12}} & \lambda_1 f_1 \\
-\lambda_2 v \, \frac{f_1}{f_{12}} & \lambda_2 f_2 \end{array} \right)
\left( \begin{array}{l} 
\!\!u\!\! \\ \!\!\chi\!\! \end{array} \right)
\label{eq:fermass}
\eea
Since $\lambda_2 >> \lambda_1$ for the light quarks we see that
the heavy quarks are approximately $u^c_2,\chi$ with masses
$\lambda_2 f_2\sim 1$ TeV, and the light (SM) quarks are $u^c_1, u$
with masses $\lambda_1 v f_2/ f_{12}$. In addition, there
is small mixing between light and heavy quarks. The mixing
between the $u^c$ fields is not physical and can be removed by
a change of basis. However, mixing
between the $SU(2)$ doublet component $u$ and the
singlet $\chi$ is significant because it alters the
couplings of up-type quarks to the $W$ and $Z$. The mixing angle is
$ \sim v f_1 /(f_2 f_{12})$, which reduces the
coupling of an up-type quark by
\bea
\delta g= -\frac12 \left({f_1 v \over f_2 f_{12}}\right)^2 \ .
\eea
For $f_1 \sim f_2 \sim 1$ TeV and $v=175$ GeV the shift in
the coupling is 1\%. A similar shift also occurs in the couplings
of neutrinos from their mixing with heavy partners.
This is problematic because
precision measurements at LEP and SLC have determined the gauge
couplings of light fermions to a precision of
$\sim 2\times 10^{-3}$ \cite{LEPSLC}.
However, we also see that it is easy to strongly suppress the mixing
by taking $f_2>f_1$. For example, taking $f_2 = 2$ TeV and
$f_1 = 1$ TeV we have
$\delta\, g \sim  10^{-3}$.
We see that the part of parameter space with $f_2>f_1$ is preferred.

We should check that taking unequal $f_i$ allows a large enough
top Yukawa coupling and does not destabilize the Higgs mass.
Diagonalizing eq.~(\ref{eq:fermass}) to leading order
in $v^2/f^2$ for the third generation we
obtain the mass of the heavy partner of the top and the
top Yukawa coupling
\bea
m_\chi = \sqrt{\lambda_1^2 f_1^2 + \lambda_2^2 f_2^2}\ , \quad
\lambda_t = \lambda_1 \lambda_2 \sqrt{{f_1^2 + f_2^2 \over \lambda_1^2 f_1^2 +
  \lambda_2^2 f_2^2}}
\eea
Happily, a wide range of $f_i$ and $\lambda_i$ can give
a heavy top quark without fine tuning of the Higgs mass. 
For example, taking $f_1=.5 $ TeV, $f_2=2$ TeV,
$\lambda_1 = \sqrt{2}$ and $\lambda_2=1/3$ we obtain the correct top
Yukawa.
To estimate the degree of fine-tuning recall that the top loop
contribution to the Higgs mass is cut off by $m_\chi$.
Thus $\delta m_h^2 \sim m_\chi^2\  \lambda_t^2 / 16 \pi^2$ which requires
no fine tuning for
$m_\chi \simeq 1$ TeV.

We now turn to computing the masses and mixings of the gauge bosons
in the full $SU(4)\times U(1)_X$ model. Transitions mediated by
the heavy $SU(4)$ gauge bosons contribute to precision electroweak
measurements leading to constraints on the $f_i$.
A useful parametrization of the non-linear sigma model fields $\Phi_i$
and $\Psi_i$ with general $f_i$ is
\bea
\Phi_1=e^{+i {\cal H}_u {f_2 \over f_1}} 
        \left( \begin{array}{l} 0  \\ 0 \\ f_1 \\0 \end{array} \right) \qquad
\Phi_2=e^{- i {\cal H}_u {f_1 \over f_2}} 
        \left( \begin{array}{l} 0  \\ 0 \\ f_2 \\0 \end{array} \right)
        \nonumber \\
\Psi_1=e^{+i {\cal H}_d {f_4 \over f_3}} 
        \left( \begin{array}{l} 0  \\ 0 \\ 0 \\f_3 \end{array} \right) \qquad
\Psi_2=e^{- i {\cal H}_d {f_3 \over f_4}}
        \left( \begin{array}{l} 0  \\ 0 \\ 0 \\f_4 \end{array} \right)
\eea
where
\bea
  {\cal H}_u =
    \left( \begin{array}{ccl} 
      \!\!\begin{array}{ll} 0 & 0 \\ 0 & 0 \end{array} 
      & \!\!h_u & \!\! \begin{array}{l}\!\!  0 \\ \!\! 0 \end{array}   \\ 
      h_u^\dagger & \!\!0 & \!\! 0 \\
      \!\! \begin{array}{cc}  0 &  0 \end{array} &\!\!0 & \!\!0 \\ 
    \end{array} \right)/f_{12}& \qquad
  {\cal H}_d =
    \left( \begin{array}{ccl} 
      \!\!\begin{array}{ll} 0 & 0 \\ 0 & 0 \end{array} 
      & \!\! \begin{array}{l}\!\!  0 \\ \!\! 0 \end{array}  & \!\!\! h_d \\ 
       \!\! \begin{array}{cc}  0 &  0 \end{array}  & \!\!0 & \!\! 0 \\
      h_d^\dagger & \!\! 0 & \!\!0 \\ 
    \end{array} \right)/ f_{34} 
\eea
Here we ignore the small contributions to masses from
small vevs for the fields $\sigma_i$.
The photon and the $Z$ are linear combinations of four
neutral gauge bosons: three gauge bosons which correspond to the
diagonal $SU(4)$ generators
$T^3=\frac{1}{2}\, diag(1,-1,0,0)$, $T^{12}=\frac{1}{2}\, diag(0,0,1,-1)$, and
$T^{15}=\frac{1}{\sqrt{8}}\, diag(-1,-1,1,1)$,
and the $U(1)_X$ gauge field $B_\mu^x$.
Two linear combinations obtain masses of order $f$
from the kinetic terms of the $\Phi_i$ and $\Psi_i$
\begin{eqnarray}
\left| \frac{g}{2} A_{\mu}^{12} + \frac{g}{2\sqrt{2}} A_{\mu}^{15} 
        - \frac{g_x}{4} B_{\mu}^x \right|^2 f_{12}^2 \nonumber\\
\left| - \frac{g}{2} A_{\mu}^{12} + \frac{g}{2\sqrt{2}} A_{\mu}^{15} 
        - \frac{g_x}{4} B_{\mu}^x \right|^2 f_{34}^2
\end{eqnarray}
where $g, g_x$ are the $SU(4)$ and $U(1)$ gauge couplings
and $f_{ij}^2=f_i^2+f_j^2$.
In the following we specialize to $f\equiv f_{12}=f_{34}$ for which the
heavy gauge boson mass matrix simplifies.
Then the two heavy eigenstates are
\begin{eqnarray}
Z''_{\mu} &=& A_{\mu}^{12}
\nonumber\\
Z'_{\mu} &=& {\sqrt{2}g A_{\mu}^{15} - g_x B_{\mu}^x \over
\sqrt{2g^2 + g_x^2}}\ ,
\nonumber
\end{eqnarray}
with masses $m_{Z''}=g f$ and $m_{Z'}=\frac{g f}{2} \sqrt{2+g_x^2/g^2} $.
The two eigenstates which remain massless at this order are
\begin{eqnarray}
W^3_\mu &=& A^3_\mu 
\nonumber\\
B_{\mu} &=& {g_x A_{\mu}^{15} + \sqrt{2}g B_{\mu}^x \over
\sqrt{2g^2 + g_x^2}}
 \ .
\end{eqnarray}
The $Z$  obtains its mass from the Higgs vevs $v = \sqrt{v_u^2 + v_d^2}$.
Ignoring mixing with the $Z'$ the mass term is
\begin{eqnarray}
v^2 \left| \frac{g}{2} W_{\mu}^3 -
           \frac{g_x}{2}\frac{1}{\sqrt{1 + g_x^2/2g^2}} B_{\mu}
        \right|^2 \ .
\end{eqnarray}
From this expression we can read off the standard model gauge couplings.
We see that the $SU(2)$ coupling of the standard model is equal to the
$SU(4)$ coupling $g$ and -- setting the coefficient of $B_\mu$ equal to
$g'/2$ -- we have
\beq
g'=g_x / \sqrt{1+\frac{g_x^2}{2g^2}} \ .
\eeq

Deviations from the standard model arise in this model at
order $v^2/f^2$ from mixing of the $Z$ with the $Z'$.
Explicitly, the mixing is determined by diagonalizing
the $Z$--$Z'$ mass matrix
\beq
\frac{g^2}{2} \left(\! \begin{array}{cc} 
v^2 (1+t^2) & 
\!-v^2 {(1-t^2)\sqrt{1+t^2} / \sqrt{2-t^2}} \\
\!-v^2 {(1-t^2)\sqrt{1+t^2} / \sqrt{2-t^2}} &
2 f^2 / (2-t^2) \end{array} \!\right)
\eeq
where $t=g'/g=\tan{\theta_W}$ and $\theta_W$ is the weak mixing angle.
Diagonalizing, we find a contribution to $\delta \rho$ from the
shift in the $Z$ mass
\beq
\delta\rho \equiv {\delta m_W^2 \over m_W^2} - {\delta m_Z^2 \over m_Z^2}
           = {v^2\over 2 f^2} (1-t^2)^2 \approx + 1.5 \cdot 10^{-3}
           \left({{\rm 2.2 TeV}\over f}\right)^2 \ .
\eeq
Given that a standard model fit predicts a $W$-mass which is lower
than the experimental value by about $1.6 \sigma$ \cite{LEPSLC},
this correction actually improves the precision electroweak
fit for $f\sim 2.2$ TeV.
Alternatively, demanding a fit that is at least as good as the standard
model implies a bound of $f\gsim 1.5$ TeV.

Another observable affected by the new gauge bosons are four-fermion
operators.  The bound on new contributions to the four-electron
operator, for example, is quite severe.  The exchange of the $Z'$
produces an operator of the size:
\begin{equation}
\frac{(1-t^2)^2}{8 f^2} {\bar e} \gamma^{\mu} e {\bar e} \gamma_{\mu} e
\end{equation}
for left-left currents. Using current bounds on this operator
we find the requirement that $f\gsim 1.5$ TeV.

\noindent {\it ii. the spectrum and direct searches:}
In the UV the standard model $SU(2)\times U(1)$ gauge group is
enlarged to $SU(4)\times U(1)$. Thus there are 12 new massive gauge
bosons with masses near a TeV. Two of them are the $Z'$ and $Z''$
discussed above. For the parameter choice
$f_1^2+f_2^2=f_3^2+f_4^2\equiv f^2$ the $Z''$ has mass $g f$ and does
not couple to standard model fermions. The $Z'$ has mass $.77 g f$
and couples to quarks and leptons.
At the Tevatron it would appear as an s-channel resonance which
decays to pairs of leptons. The limit on the mass
of such a $Z'$ from CDF \cite{CDFzprime}
is in the 700-800 GeV range, implying a bound $f \gsim$ 1 TeV.
The off-diagonal $SU(4)$ gauge bosons and their masses are
\bea
    \left( \begin{array}{cccc} 
       &  & Y^0 & Y'^0 \\ 
       &  & X^- & X'^- \\ 
      \bar Y^0 & X^+ &  & Y''^0  \\
      \bar Y'^{0} & X'^+  & \bar Y''^0 &  \\ 
    \end{array} \right) = 
     \left( \begin{array}{cccc} 
       &  & .5  & .5 \\ 
       &  & .5 & .5 \\ 
     .5  & .5  &  & 1 \\
     .5  & .5 & 1 &  \\ 
    \end{array} \right) \ g f
\eea
The $Y''^0$ only couples to the heavy fermions and is therefore
extremely difficult to detect. All others couple to one light and
one heavy fermion. They can be produced in association
with a heavy fermion or else appear in t-channel diagrams.

There are two vector-like heavy quarks of charge $2/3$
for each generation. As we discussed above, one of them mixes
with up-type quarks and can be produced singly in t-channel
$W$ exchange. The LHC reach in this channel can be as large as
several TeV \cite{taohan}. The masses of these quarks are not
completely determined because they depend on unknown 
Yukawa couplings. But flavor constraints suggest that their
masses are generation independent and since naturalness requires
a partner for the top quark below $\sim$ 2 TeV we expect
at least one set of these quarks to be visible at the LHC.
In terms of model parameters, the new quark masses are
$\sqrt{\lambda_1^2 f_1^2 + \lambda_2^2 f_2^2}$ and  $\lambda_3 f_3$.

In addition, there are also vector-like heavy leptons which
mix with the neutrinos. These fermions are impossible to
discover directly but their existence might be inferred by
missing energy signals or through their mixing with the light neutrinos
in precision data.
Note also that this mixing reduces neutral
currents of neutrinos more than charged currents
which might help to explain the NuTeV anomaly \cite{takeuchi}
and the slightly reduced invisible width of the $Z$.

The scalar spectrum consists of the two Higgs doublets with
masses near the weak scale, three complex neutral fields 
$\sigma_i$ with masses $\sim \kappa f$, and two real scalar fields 
$\eta_i$ with masses of order the weak scale. The $\eta_i$ only couple
to heavy fields and in the $b$-terms of the Higgs potential, thus
they are very difficult to detect despite their relatively low masses.

\noindent {\it iii. flavor:}
The $SU(4)$ model has more Yukawa matrices than the
standard model. In general these additional sources of flavor violation
lead to flavor changing neutral currents. The easiest way to
suppress the flavor violation is to assume that the new Yukawa
matrices are proportional to the unit matrix. This assumption
imposes constraints on the UV completion of the theory,
but it is ``technically natural'' in the effective theory (loops
in the effective theory only generate small corrections).
Note that this is similar to the assumption of universal soft
masses in supersymmetry.

For example, the quark Yukawa couplings are then 
\beq
{\cal L}_{quarks} = (\lambda^{u}\, u_1^{c}\, \Phi_1^\dagger  
        + I_1\; u_2^{c}\, \Phi_2^\dagger 
                + I_2\, u_3^{c}\, \Psi_1^\dagger) Q 
                +\: \lambda^d\; d^c \, \Phi_1 \Psi_1 \Psi_2 Q
\label{eq:quarkyukagain}
\eeq
where $\lambda^{u}$ and $\lambda^{d}$ are similar to the usual
standard model Yukawa couplings and $I_1$ and $I_2$ are approximately
proportional to the unit matrix (in flavor space). To see that
these couplings do not contain dangerous flavor violation, we
go to a new basis in which $\lambda^{u}$ is diagonal. 
It is convenient to rotate all four components of $Q$ in the same way.
$I_1$ and $I_2$ remain unit matrices if $u_2^{c}$ and
$u_3^{c}$ are rotated appropriately. Finally, we also diagonalize
$\lambda^{d}$ with a bi-unitary transformation, but this time
we transform only the down-type quarks in $Q$.
In the new basis all Yukawa couplings are diagonal; flavor violation
resides only in the gauge couplings of $W$, $X$, $Y$
to quarks and in couplings of multiple Higgses to quarks which arise
from expanding out the down Yukawa operator. The latter
couplings are small and only appear in loops. The former are also
easily shown to be harmless. There are the usual $W$ couplings proportional
to the CKM matrix in addition to new couplings between one down-type quark,
one heavy vector-like up-type quark and the heavy gauge bosons
$X$ and $Y$. These couplings allow box diagrams and penguins which
are similar to corresponding standard model diagrams with $W$s replaced
by $X$s or $Y$s. The resulting flavor changing neutral currents are
suppressed relative to the  standard model ones by the large masses
of the heavy gauge bosons $m_W^2/m_X^2$ and can be ignored.

A similar analysis of the lepton sector shows that there is also
no dangerous lepton flavor violation in the low energy theory. 
Of course, the theory may also contain direct flavor violating
four Fermi operators suppressed by the cut-off $\Lambda$. Such
operators are constrained by $K-\overline{K}$ mixing and CP
violation. The experimental bounds on such operators therefore
imply constraints on the unknown UV-theory above 10 TeV.

Small neutrino masses can be obtained by including a higher
dimensional lepton number violating operator $(\Phi^\dagger L)^2$
with a small coefficient in the effective theory. This operator
might arise from a generalization of the see-saw mechanism in
the UV completion: supermassive right handed neutrinos coupled
to the operator which interpolates $(\Phi^\dagger L)$ in
the UV theory.

\section{Discussion}
\label{disc}

We have seen that, in a little Higgs model,
embedding $SU(2)_{weak}$ into a simple group
(such as $SU(3)$) is enough to cancel one-loop
quadratic divergences from gauge and
(perturbatively coupled) fermion loops.
The $SU(3)$ model only lacks a quartic.
One possibility is to
simply ignore the relatively insignificant fine-tuning from
the Higgs couplings and add the quartic by hand in ``component''
fields ({\it i.e.}, not the full $\Phi$).  This coupling need not be very
large as an additional contribution to the quartic would come from
the log-divergent and finite contributions to the effective potential
from the top sector below the scale $f$.  At worst, this introduces
of order 10\% fine-tuning.

There are at least three different possibilities for ultra-violet completions
to our models.  One is a linear sigma model with supersymmetry protecting
the scalar masses above the multi-TeV scale.  The $SU(3)$ model would work
well in this case as the quartic would be provided by the $D$-term and the
Higgs would remain massless at tree-level as long as there are more than
two triplets.  In addition, the group is simple enough to embed
in a unifying theory:
$SU(3+n)_{color} \times SU(2+n)_{weak} \times U(1)$ gives
coupling constant predictions extremely close to those of the MSSM, when
charges are normalized to embed into $SU(5+2n)$, and matter is in complete
representations except for a split fundamental and
anti-fundamental \cite{ann}.

If one wishes to complete the theory above $4\pi f$ with a strongly coupled
theory, the coset space we've used would require something different than
a QCD-like model.  For example, a gauged $SU(7)$
with four fundamentals, one anti-fundamental and one anti-symmetric tensor
produces the symmetry-breaking pattern $SU(4)\rightarrow SU(3)$ assuming the
fundamentals condense with the anti-fundamental.
Fermion and quartic interactions
would require additional dynamics
as in extended-technicolor \cite{etc}.

A more interesting possibility would be a linear sigma model completing
into another little Higgs theory at a higher scale, $F\sim 10 TeV$.
This may be possible in the $SU(3)$ theory if the ``fundamental'' quartic
added need not be too large.  One example would be
$[SU(7)/SO(7)]^2$ with $SU(3)$ subgroups gauged similar to
the ``littlest Higgs'' \cite{Arkani-Hamed:2002qy}.
This could produce two light
$SU(3)$ triplets with fermion couplings causing vacuum misalignment
and $SU(3)$-breaking at a scale $f \equiv F/4\pi$.  If a model of this type
works, it would provide a weakly coupled theory of electroweak symmetry
breaking valid up to $>$100 TeV.

\section{Acknowledgements}

We thank Nima Arkani-Hamed, Andy Cohen and Sekhar Chivukula
for numerous helpful discussions.
DK thanks the visitor program of the particle theory group at the 
Boston University Physics Department for their hospitality,
and we thank the Aspen Center for Physics for providing a stimulating
environment during the early phase of this work.
MS is supported by the DOE grant DE-FG02-90ER-40560 and the
Outstanding Junior Investigator Award DE-FG02-91ER40676.

\end{document}